\documentclass[twoside]{article}
\usepackage{graphicx,amssymb,mathrsfs,amsmath}
\catcode`@=11
\long\def\@makefntext#1{\noindent #1}
\newskip\tabcentering \tabcentering=1000pt plus 1000pt minus 1000pt
\def\REF#1{\par\hangindent\parindent\indent\llap{#1\enspace}\ignorespaces} %reference format
\def\MCH#1#2{\setbox0=\hbox{\raise#1\hbox{#2}}\smash{\box0}}% move char

\def\@evenfoot{}\def\@oddfoot{}

\def\@evenhead{\hbox to\textwidth{\footnotesize\rm\thepage \hfill
{\it Jia Xu, Yong Yao}}} % authors name

\def\@oddhead{\hbox to \textwidth{\footnotesize{\it
Majorization Order, Termination} \hfill\thepage}}% abbreviate title

\catcode`@=12

\def\bc{\begin{center}}
\def\ec{\end{center}}
\def\no{\noindent}
\def\hang{\hangindent\parindent}
\def\textindent#1{\indent\llap{\qquad #1\ \ \enspace}\ignorespaces}
\def\ref{\par\hang\textindent}

\title{The Majorization Order on Monomials and \\
Termination of the Successive Difference Substitutions\thanks{The work was supported by the National
Natural Science Foundation of China(11001228), and the Fundamental
Research Funds for the Central Universities, Southwest University
for Nationalities (12NZYTH04).}}

\author{Jia Xu\\
College of Computer Science and Technology, \\
Southwest University for Nationalities, \\
Chengdu, Sichuan 610041, PR China \\
E-mail: {\tt j.jia.xu@gmail.com} \\ \\
Yong Yao\\
Chengdu Institute of Computer Applications, \\
Chinese Academy of Sciences, \\
Chengdu, Sichuan 610041, PR China \\
E-mail: {\tt yaoyong@casit.ac.cn}
}
\begin{document}
\maketitle
\begin{abstract}
We introduce a majorization order on monomials. With the help of this order,
we derive a necessary condition on the positive termination of a general
successive difference substitution algorithm (KSDS) for an input form $f$.
\end{abstract}

%{\small \bf Keywords:} Successive difference substitution
%algorithm, majorization order on monomials, termination, positive
%semi-definite form

\section{Introduction}

The first successive difference substitution algorithm (SDS) based on the
matrix
$$
A_{n}=
\begin{pmatrix}
1 & 1 & \cdots & 1 \\
& 1 & \ddots & \vdots \\
&  &\ddots &1 \\
0 &  &  & 1
\end{pmatrix}$$
originates from proving homogeneous symmetric inequalities. It was
developed by L.~Yang in [1], [2] and [3], and improved
subsequently in~[4] and~[5].  In particular, Y.~Yao established a
new successive difference substitution algorithm based on the matrix
$$
G_{n}=
\begin{pmatrix}
1 & \frac{1}{2} & \cdots & \frac{1}{n} \\
& \frac{1}{2} & \ddots & \vdots \\
& &\ddots &\frac{1}{n} \\
0 &  &  & \frac{1}{n}
\end{pmatrix}
.
$$
His method is named as NEWTSDS, which has many interesting
properties (see [5]).  These results illustrate that SDS may be  an
effective tool for solving many problems in real algebra.

However, it is still very hard to find necessary and/or sufficient
conditions on the termination of SDS and NEWTSDS. In this paper, we
will study the termination of a general successive difference
substitution algorithm (KSDS) by the majorization order on
monomials. Our main result is as follows:

\medskip \noindent
\textbf{Main result} A necessary condition of positively terminating
of KSDS for an input $f$ is that, for an arbitrary ordering of
variables, every monomial of $f$ with negative coefficient is
majorized by at least one monomial of $f$ with positive coefficient.

\medskip
The paper is organized as follows. In Section~2, we introduce KSDS and present some
background materials. In Section~3, we discuss necessary
conditions on the termination of KSDS using the majorization order
on monomials. The future research directions are outlined in
Section~4.

\section{General successive difference
substitution - KSDS}

Let ${\bf \alpha}=(\alpha_1,\ldots,\alpha_n)\in \mathbb{N}^n$. We
set~$|\alpha|=\alpha_1+\cdots+\alpha_n$. A form (i.e., a homogeneous
polynomial)~$f$ of degree $d$ can be written as
\begin{eqnarray}
f(x_1,\ldots,x_n)=\sum\limits_{|\alpha|=d}C_{\alpha}x^{\alpha_1}_1
\cdots x^{\alpha_n}_n
=\sum\limits_{|\alpha|=d}C_{\alpha}X^{\alpha},\ C_{\alpha}\in
{\mathbb R}.
\end{eqnarray}
The next definition is given in~[5].

\medskip \noindent
{\bf Definition 2.1.} A form $f$ is said to be {\em trivially
positive} if the coefficient~$C_{\alpha}$ of every monomial
$X^{\alpha}$ is nonnegative. It is said to be
{\em trivially negative} if $f(1,1,\ldots,1)<0$ (i.e., the sum
of coefficients of $f$ is less than zero).

\medskip \noindent
{\bf Definition 2.2.} A form $f(x_1,\ldots,x_n)\in {\mathbb
R}[x_1,\ldots,x_n]$ is {\em positive semi-definite} on ${\mathbb R}^n_+$
if it satisfies
$$\forall\ (x_1,\ldots,x_n) \in {\mathbb R}^n_+, \, f(x_1,\ldots,x_n)\geq 0,$$
where ${\mathbb R}^n_+=\{(x_1,\ldots,x_n) \, \mid \, x_1\geq 0,\ldots, x_n\geq
0\}$. We denote by PSD the set of all the positive semi-definite
forms on ${\mathbb R}^n_+$. Furthermore, a positive semi-definite
form $f$ is said to be {\em positive definite} on ${\mathbb R}^n_+$ if $f>
0$ for~$(x_1,\ldots,x_n)\neq (0,\ldots,0)$. The set
of all the positive definite forms is denoted by PD.

\medskip
There are two obvious results describing the relation
between trivially positive (negative) and PSD:
\begin{enumerate}
\item If a form
$f$ is trivially positive, then $f\in$ PSD.
\item If a form $f$
is trivially negative, then $f \notin$ PSD.
\end{enumerate}

Given positive real numbers~$q_1, \ldots, q_n$, we consider the
matrix
\begin{eqnarray}\label{eqa2}
K_n =
\begin{pmatrix}
q_1 & q_2 & \cdots & q_n \\
 & q_2  & \ddots & \vdots \\
 &  &\ddots &q_n \\
0  &   &  & q_n
\end{pmatrix}.
\end{eqnarray}

Notice that $K_n=A_n$ if $q_1=q_2=\cdots=q_n=1$, and
that~$K_n=G_n$ if~$q_1=1,$ $q_2=\frac{1}{2},$ \ldots, $q_i=\frac{1}{i},$ \ldots, $q_n=\frac{1}{n}$.
So~$K_n$ is a general form of the matrices including~$A_n$ and~$G_n$.

Suppose that $S_n$ is a symmetric group of degree $n$. For
$\sigma\in S_n$, let $P_{\sigma}$ be an $n \times n$ permutation
matrix corresponding to $\sigma$. For example, suppose that
$\sigma=(1)(23)$ is a permutation. Then it corresponds to the matrix
$$P_{(1)(23)}=
\left ( \begin{array}{ccc}
1&0&0\\
0&0&1\\
0&1&0
\end{array}
\right),$$ in which  the second and third rows are permuted from the
identity matrix.

Using the notation in~[5], we introduce a few terminologies.

\medskip \noindent
{\bf Definition 2.3.} The $n\times n$ matrix
$B_{\sigma}$ with~$\sigma \in S_n$ is defined by
\begin{eqnarray*}
B_{\sigma}=P_{\sigma}K_n.
\end{eqnarray*}

\medskip
As an example, let us consider again~$\sigma=(1)(23)$. Then
$$B_{(1)(23)}=P_{(1)(23)}K_3= \left (
\begin{array}{ccc}
q_{1}&q_{2}&q_{3}\\
0&0&q_{3}\\
0&q_{2}&q_{3}
\end{array}
\right).
$$

\medskip \noindent
{\bf Definition 2.4.} Let~$f\in {\Bbb
R}[x_1,\ldots,x_n]$ and~$X=(x_1,\ldots,x_n)^T$.  Define
$${\rm SDS}_{K}(f)=\bigcup \limits_{\sigma\in S_n}f(B_{\sigma}X).$$
The set ${\rm SDS}_{K}(f)$ is called the set of difference
substitution for~$f$ based on the matrix $K_n$.

It is easy to show the following equivalence relations (see [5])
$$f\in {\rm PSD} \Longleftrightarrow {\rm
SDS_K}(f)\subset {\rm PSD} \quad \text{and} \quad f\notin \ {\rm PSD}
\Longleftrightarrow \exists g\in {\rm SDS_K}(f),g\notin \  {\rm
PSD}.$$

Repeatedly using the above two equivalence relations and Definition
2.1, we have an algorithm for testing positive
semi-definite of polynomials, which is called the {\em successive
difference substitution algorithm based on the matrix $K_n$} (KSDS) in~[5].
%Since $K_n$ is a general form of some matrices just like $A_n$ or
%$G_n$, thus we call the algorithm KSDS a general successive
%difference substitution algorithm. A crucial fact is that NEWTSDS and SDS are all special cases of KSDS.\\
\\

\begin{tabular}{l}
\hline
\indent{ \textbf{Algorithm KSDS} }\\
\hline
Input: A form $f\in {\Bbb Q}[x_1,x_2,\ldots,x_n]$.\\
Output: \lq\lq $f\in {\rm PSD}$ \rq\rq or \lq\lq $f\notin {\rm PSD}$ \rq\rq.\\
K1: Let $F=\{f \}$.\\
K2: Compute $T:=\bigcup \limits_{g\in F} {\rm SDS}_{K}(g)$,
Temp:=$T\setminus \{\hbox{ tivially positive polynomials of
$T$}\}$.\\

\qquad \quad  K21: If Temp=$\emptyset$, then return \lq\lq $f\in {\rm
PSD}$\rq\rq.\\

\qquad \quad  K22: Else if there are trivially negative forms in
Temp then return \lq\lq $f\notin {\rm PSD}$ \rq\rq.\\

\qquad \quad  K23: Else let $F={\rm Temp}$ and go to step K2.\\

\hline
\end{tabular}\\

There is a fundamental question on the algorithm KSDS. Namely, under
what conditions does the algorithm terminate? This question is very
hard to solve. Quite recently, Yang and Yao ([4], [5]) obtained some
results about the termination of SDS and NEWTSDS. Their results lead
to the following definition.

\medskip \noindent
{\bf Definition 2.5.} The algorithm KSDS is positively
terminating if the output is "$f\in {\rm PSD}$" for the input $f$. The
algorithm KSDS is negatively terminating if the output is "$f\notin
{\rm PSD}$"  for the input $f$. Otherwise, KSDS is not terminating for~$f$.

\medskip

According to Definition 2.5, it is easy to get the following assertions.

\medskip \noindent
{\bf Lemma 2.1}
\begin{enumerate}
\item The algorithm KSDS is positively
terminating for an input $f$ if and only if there exists a positive integer $m$
such that all of the coefficients of the polynomial
$$f(B_{\sigma_1}B_{\sigma_2}\cdots B_{\sigma_m}X),\ \forall \sigma_i
\in S_n,\ i=1,\ldots, m$$ are positive.

\item The algorithm KSDS is negatively terminating if and only if there exist $m$
permutations $\sigma_1, \ldots, \sigma_m \in S_n$ such that
$$f\left(B_{\sigma_1}B_{\sigma_2}\cdots
B_{\sigma_m}(1,1,\cdots,1)^T \right)<0.$$
\end{enumerate}

\section{Majorization order on monomials and the main result}

Given two monomials
$$X^{\alpha}=x_1^{\alpha_1}\cdots x_n^{\alpha_n} \quad \text{and} \quad
X^{\beta}=x_1^{\beta_1}\cdots x_n^{\beta_n}$$
with~$|\alpha|=|\beta|,$ we cannot order them unless
some further conditions are imposed. For example, let $
\alpha=(3,1,1),\beta=(2,1,2)$ and $x_1\geq x_2 \geq x_3 \geq 0$,
then we have
$$x_1^3x_2x_3- x_1^2x_2x_3^2=x_1^2x_2x_3(x_1-x_3)\geq 0.$$

This example inspires us to use a majorization order on monomials
for our analysis of the termination of KSDS.

Before that, we first introduce the majorization between two vectors
given in [6] and [7].

\medskip \noindent
{\bf Definition 3.1.} Let
${\alpha}=(\alpha_1,\ldots,\alpha_n)$ and~${\beta}=(\beta_1,
\ldots,\beta_n)$, where~$\alpha,\beta\in {\Bbb R^n_+}$ with~$|\alpha|=|\beta|$.
If
\begin{eqnarray}
\sum^k_{i=1}\alpha_i\geq \sum^k_{i=1}\beta_i \quad \mbox{for all~$k \in \{ 1,\cdots,n-1\},$}
\end{eqnarray}
then we say that~${\alpha}$ majorizes~${\beta}$, which is denoted as~${\alpha}\succeq {\beta}$.

\medskip
Note that~\lq\lq $\succeq$\rq\rq is a partial order.
With the help of Definition 3.1, we construct the definition of
majorization order on monomials.

\medskip \noindent
{\bf Definition 3.2} (Majorization order on monomials)\quad Let
$X^{\alpha}$ and~$X^{\beta}$ be two monomials
with~$|\alpha|=|\beta|$. Suppose that $\sigma$ is a permutation on
the set $\{1,2,\ldots,n\}$. If
$$(\alpha_{\sigma(1)},\ldots,\alpha_{\sigma(n)})\succeq
 (\beta_{\sigma(1)},\ldots,\beta_{\sigma(n)}) \quad \hbox{or, briefly,} \quad
   \alpha_{\sigma}\succeq \beta_{\sigma},$$
 then we say that $X^{\alpha}$  majorizes $X^{\beta}$ with respect to
 the permutation
 $\sigma$,
 %in the ordering
 %of variables $x_{\sigma(1)},
%x_{\sigma(2)}, \cdots , x_{\sigma(n)}$,
which is denoted as
$(X^{\alpha})_{\sigma} \succeq (X^{\beta})_{\sigma}$ or
$X^{{\alpha}_{\sigma}}_{\sigma} \succeq
X^{{\beta}_{\sigma}}_{\sigma}$ .

\medskip
Our definition of the majorization order on monomials evolves from
the definition of the majorization on symmetric
polynomials given in~[6] and~[7].

We need a few comments on the notation.
Note that $X^{\alpha}$, $(X^{\alpha})_{\sigma}$ and
$X_{\sigma}^{\alpha_{\sigma}}$ stand for the same monomial. Furthermore, there is
$$(X^{\alpha})_{\sigma} \succeq (X^{\beta})_{\sigma}
\Longleftrightarrow (X_{\sigma}^{\alpha_{\sigma}})_I \succeq
 (X_{\sigma}^{\beta_{\sigma}})_I,$$
where $I$ is the identical permutation and can be omitted.
For example
\begin{eqnarray*}
(x_1^3x_2^4x_3)_{(21)(3)}\succeq (x_1^4x_2^2x_3^2)_{(21)(3)}
\Longleftrightarrow x_2^4x_1^3x_3\succeq x_2^2x_1^4x_3^2
\Longleftrightarrow (4,3,1)\succeq (2,4,2).
\end{eqnarray*}

It is easy to see that, with respect to the permutation
$\sigma=(1)(2)(3)$, the monomials $x_1^3x_2^4x_3$ and
$x_1^4x_2^2x_3^2$ do not majorize each other. So the majorization
order on monomials is a partial order. Moreover, the
following three basic properties hold.

\medskip \noindent
{\bf Lemma 3.1.} For a given
permutation $\sigma\in S_n$ and~$\alpha, \beta, \gamma \in \mathbb N^n$ with~$|\alpha|=|\beta|=|\gamma|$,
we have
\begin{enumerate}
\item $(X^{\alpha})_{\sigma}\succeq (X^{\alpha})_{\sigma}.$

\item $(X^{\alpha})_{\sigma}\succeq (X^{\beta})_{\sigma} \wedge
(X^{\beta})_{\sigma}\succeq (X^{\alpha})_{\sigma} \Longrightarrow
X^{\alpha}=X^{\beta}.$

\item $(X^{\alpha})_{\sigma}\succeq (X^{\beta})_{\sigma} \wedge
(X^{\beta})_{\sigma} \succeq (X^{\gamma})_{\sigma}
 \Longrightarrow (X^{\alpha})_{\sigma} \succeq
 (X^{\gamma})_{\sigma}$.
\end{enumerate}
\begin{proof}
Straightforward. \quad $\blacksquare$
\end{proof}

\medskip \noindent
{\bf Lemma 3.2.}  Let $\sigma \in S_n$ be a given permutation. For
the monomial $X^{\alpha}$ and $X^{\beta}$ with $|\alpha|=|\beta|$,
we have $X^{\alpha}\geq X^{\beta}$ under the condition
$x_{\sigma(1)}\geq \cdots \geq x_{\sigma(n)}\geq 0 $ if and only if
$(X^{\alpha})_{\sigma}\succeq (X^{\beta})_{\sigma}.$

\begin{proof}
$\Rightarrow$: Let $x_{\sigma(1)}=\cdots=x_{\sigma(j)}=2$, and let
$x_{\sigma(j+1)}=\cdots=x_{\sigma(n)}=1$. Then
$$2^{\alpha_{\sigma(1)}+\alpha_{\sigma(2)}+\cdots+\alpha_{\sigma(j)}}
\geq
2^{\beta_{\sigma(1)}+\beta_{\sigma(2)}+\cdots+\alpha_{\sigma(j)}}.$$
Thus
$${\alpha_{\sigma(1)}+\alpha_{\sigma(2)}+\cdots+\alpha_{\sigma(j)}}\geq
{\beta_{\sigma(1)}+\beta_{\sigma(2)}+\cdots+\alpha_{\sigma(j)}}.$$
Let $j=1,2,\cdots,n-1$ successively, then we immediately have
$$(\alpha_{\sigma(1)},\cdots,\alpha_{\sigma(n)})\succeq
 (\beta_{\sigma(1)},\cdots,\beta_{\sigma(n)}).$$

$\Leftarrow$: It is trivial if $x_i=0, i=1,\ldots,n$. So we assume
that $x_i\neq 0$ for all $i=1,\ldots, n$. Then
$$\frac{X^{\alpha}}{X^{\beta}}=\prod_{i=1}^{n-1}
\left(\frac{x_{\sigma(i)}}{x_{\sigma(i+1)}}\right)
^{\sum_{j=1}^i(\alpha_{\sigma(j)}-\beta_{\sigma(j)})}\geq 1.$$ \quad
$\blacksquare$
\end{proof}

\medskip \noindent
{\bf Lemma 3.3.} Let $M=(p_{ij})$ be an $n \times n$ matrix, in
which $p_{ij}>0$ if $i\leq j$  else $p_{ij}=0$. For a monomial
$x_1^{\alpha_1}x_2^{\alpha_2}\cdots x_n^{\alpha_n}$, consider linear
substitution $(x_1,\ldots,x_n)^T=M(t_1,\ldots,t_n)^T$, namely,
$$
\begin{array}{l}
(p_{11}t_{1}+p_{12}t_2+\cdots+p_{1n}t_{n})^{\alpha_1}(p_{22}t_2+\cdots+p_{2n}t_{n})^{\alpha_2}
\cdots (p_{nn}t_{n})^{\alpha_n}\\
=\sum\limits_{|(j_1,\ldots,j_n)|=|\alpha|}
C_{(j_1,\ldots,j_n)}t_1^{j_1}t_2^{j_2}\cdots t_n^{j_n}.
\end{array}
$$
Then
$$
C_{(j_1,\ldots,j_n)}\neq 0\Longleftrightarrow (t_1^{\alpha_1}\cdots
t_n^{\alpha_n})_I\succeq (t_1^{j_1}\cdots t_n^{j_n})_I.
$$
\begin{proof}
$\Rightarrow$: Consider the expansion
$$
\sum\limits_{|(j_1,\ldots,j_n)|=|\alpha|}
C_{(j_1,\ldots,j_n)}t_1^{j_1}t_2^{j_2}\cdots t_n^{j_n}.
$$
If $C_{(j_1,\cdots,j_n)}\neq 0$, then we have the following results:

The term~$t_1^{j_1}$ can be obtained by expanding
$(p_{11}t_{1}+p_{12}t_2+\cdots+p_{1n}t_{n})^{\alpha_1}$. It follows
that $j_1\leq \alpha_1$. Analogously, $t_2^{j_2}$ can  be obtained
by expanding $(p_{11}t_{1}+p_{12}t_2+\cdots+p_{1n}t_{n})^{\alpha_1}$
or $(p_{22}t_2+\cdots+p_{2n}t_{n})^{\alpha_2}$ and therefore
$j_2\leq (\alpha_1-j_1)+{\alpha_2}$, namely, $j_1+j_2\leq
\alpha_1+{\alpha_2}$. By the same token, we have
$$(j_1,\ldots,j_n)\preceq (\alpha_1,\ldots,\alpha_n).$$
Namely $(t_1^{\alpha_1}\cdots t_n^{\alpha_n})_I\succeq
(t_1^{j_1}\cdots t_n^{j_n})_I$.

$\Leftarrow$: It is easy to see that the converse implications are
also true. \quad $\blacksquare$
\end{proof}

\medskip \noindent
{\bf Theorem 1.} Suppose that
\begin{equation}
f(x_1,\ldots,x_n)=\sum\limits_{|\alpha|=d}C_{\alpha}x^{\alpha_1}_1
\cdots x^{\alpha_n}_n=\sum\limits_{|\alpha|=d}
C_{\alpha}X^{\alpha}, \quad \text{where $C_{\alpha}\neq 0.$}
\end{equation}
is a homogeneous polynomial of degree $d$ in ${\Bbb
Q}[x_1,\ldots,x_n]$. For a term $C_{\lambda}X^{\lambda}$ of $f$, if
the monomial~$X^\lambda$ is not majorized by any other monomial of
$f$ with respect to $\sigma\in S_n$ then the coefficient of the
monomial $(X_{\sigma})^{\lambda}$ of $f(B_{\sigma}K^{m-1}_nX)$ is
$$\left(q_{\sigma(1)}^{\lambda_1}\cdots
q_{\sigma(n)}^{\lambda_n}\right)^m C_{\lambda}.$$

\begin{proof}
According to (2), we know that $K^m_n$ is an upper triangular
matrix and the diagonal elements are $q_1^m,\ldots,q^m_n$. Let
$$
K^m_n=
\begin{pmatrix}
q^m_{1}& p_{12}& \cdots & p_{1n} \\
 & q^m_{2}&\dots & p_{2n} \\
 &  &\ddots &\vdots \\
0&  &  &q^m_{n}
\end{pmatrix}
, \quad \text{where~$(p_{ij}>0,\ 1\leq i<j\leq n).$}
$$
Let
\begin{eqnarray}
X'=\left(
\begin{array}{c}
x'_1 \\
x'_2 \\
\vdots \\
x'_n
\end{array}
\right)=K^m_nX=\left(
\begin{array}{r}
q^m_{1}x_1+ p_{12}x_2+\cdots + p_{1n}x_n \\
q^m_{2}x_2+ \dots+ p_{2n}x_n \\
\cdots \\
q^m_{n}x_n
\end{array}
\right).
\end{eqnarray}
By Definition 2.3 and (4), we have the following result.
\begin{eqnarray*}
f(B_{\sigma}K^{m-1}_n X)&=&f(P_{\sigma} K^m_n X)=f(P_{\sigma}X')\\
&=&f(x'_{\sigma(1)},\ldots,x'_{\sigma(n)})
=\sum\limits_{|\alpha|=d}C_{\alpha}(X'_\sigma)^{\alpha}.
\end{eqnarray*}

Notice that the monomial $X^{\lambda}$ is not majorized by any other
monomial of $f$ with respect to $\sigma$. By Lemma 3.1 and Lemma
3.3, the monomial $(X_{\sigma})^{\lambda}$ of
$f(B_{\sigma}K^{m-1}_nX)$ is only generated by expanding
$(X'_{\sigma})^{\lambda}$. By (5), we get that the coefficient of
$(X_{\sigma})^{\lambda}$ is $\left(q_{\sigma(1)}^{\lambda_1}\cdots
q_{\sigma(n)}^{\lambda_n}\right)^m C_{\lambda}$. \quad
$\blacksquare$
\end{proof}

By Lemma 2.1 and Theorem 1, we immediately have the following main
result.

\medskip \noindent
\textbf{Theorem 2.} A necessary condition of positively terminating
of KSDS for an input form $f$ is that, for an arbitrary ordering of
variables, every monomial of $f$ with a negative coefficient is
majorized by at least one monomial of $f$ with a positive coefficient.
\begin{proof} We argue by contradiction.
Suppose that there is a term $C_{\lambda} X^{\lambda}$
($C_{\lambda}<0$) of $f$, in which $X^{\lambda}$ is not majorized by
any other monomial of $f$ with respect to $\sigma$. Then, by theorem
1, the coefficient of $X_{\sigma}^{\lambda}$ is always negative
after expanding the polynomial $f(B_{\sigma}K^{m-1}_nX)$. This is
a contradiction with Lemma~2.1. $\blacksquare$
\end{proof}

\medskip
For example, let us consider the cyclic polynomial
$$f=x_1^4x_2^2-x_1^3x_2x_3^2+x_2^4x_3^2-x_1^2x_2^3x_3+x_1^2x_3^4-x_1x_2^2x_3^3.$$
Note that the monomial $x_1^3x_2x_3^2$ in~$f$ has a negative coefficient,
which is not majorized by any other monomials $x_1^4x_2^2,\ x_2^4x_3^2,\
x_1^2x_3^4$ in~$f$ with positive coefficients
 in the ordering $x_1, x_3, x_2$. Choose the following matrix $A_3$, and let
the permutation $\sigma=(1)(23)$.
\begin{eqnarray*}
A_{3}= \left(
\begin{array}{ccc}
1 & 1 & 1 \\
0 & 1 & 1 \\
0 &  0 & 1
\end{array}
\right),\quad P_{(1)(23)}= \left(
\begin{array}{ccc}
1 & 0& 0  \\
0 & 0 & 1 \\
0 & 1 & 0
\end{array}
\right).
\end{eqnarray*}

Expanding the polynomial $f(P_{(1)(23)}A_3^m X)$,  we see that the coefficient of $x_1^3x_2^2x_3$ is always $-1$ by Theorem~1.
 So SDS (based on $A_3$) is not positively terminating for the input~$f$.
By other methods, we can prove that $\forall\ X\in {\Bbb R^3_+}, \, f\geq
0$. So SDS is not negatively terminating either.

On the other hand, using Jordan normal form, we can compute $P_{(1)(23)}A_3^m $
$$P_{(1)(23)}A_3^m=
\left(
\begin{array}{ccc}
1 & m& m(m-1)/2  \\
0 & 0 & 1 \\
0 & 1 & m
\end{array}
\right).
$$
The coefficient of $x_1^3x_2^2x_3$ is still $-1$ by expanding
$f(P_{(1)(23)}A_3^m X)$. Thus, the results obtained by the
above two methods are compatible.

\section{Conclusion}

There are many interesting questions arising from the family of successive difference
substitutions. For example, what is a necessary and
sufficient condition for the positive termination of the algorithm KSDS?
What is a necessary and sufficient condition for the negative
termination of KSDS? Some research directions are listed below:
\begin{enumerate}
\item Yang and Yao  proved that a necessary and
sufficient condition on the negative termination of SDS and NEWTSDS is
$f\notin {\rm PSD}$ (see [4] and [5]). So  we put forward a conjecture
for KSDS.

\medskip \noindent
\textbf{Conjecture.} The algorithm KSDS is negatively terminating
if and only if $f\notin {\rm PSD}$.

\item For the positive termination of NEWTSDS, Yao has proved the
following result in [5].

\medskip \noindent
{\bf Theorem 3.} Let $f(X)\in {\Bbb
R}[x_1,\cdots,x_n]$. If $(\forall X\in {\Bbb R^n_+},X\neq 0)\ f(X)>
0$, then there exists $m>0$ such that the coefficients of
$$f(B_{\sigma_1}B_{\sigma_2}\cdots B_{\sigma_m}X),\ \forall \sigma_i\in
S_n\ ( B_{\sigma_i}=P_{\sigma_i}G_n )$$ are all positive.

\medskip
In other words, NEWTSDS is positively terminating for a form in PD.
However, it appears more difficult to study the positive termination of
KSDS.
\end{enumerate}

\no \vskip0.2in
\no {\bf References}
\vskip0.1in

\footnotesize

\REF{[1]} L. Yang, Solving Harder Problems with Lesser Mathematics.
Proceedings of the 10th Asian Technology Conference in Mathematics,
ATCM Inc, 2005, 37-46.

\REF{[2]} L. Yang, Difference Substitution and Automated Inequality
Proving, Journal of Guangzhou University (Natural Science Edition),
2006, 5(2), 1-7.

\REF{[3]} L. Yang, B. Xia, Automated Proving and Discovering on
Inequalities (in Chinese). Beijing: Science Press, 2008, 22, 174.

\REF{[4]} L. Yang, Y. Yao, Difference substitution Matrices and
Decision on Nonnegativity of Polynomials (in Chinese), Journal of
Systems Sciences and Mathematical Sciences, 2009, 29(9), 1169-1177.

\REF{[5]} Y. Yao, Successive Difference Substitution Based on Column
Stochastic Matrix and Mechanical Decision for Positive
 Semi-definite Forms (in Chinese), Sci Sin Math, 2010, 40(3),
 251-264. \quad(Also see http://arxiv.org/abs/0904.4030v3)

\REF{[6]} G.H. Hardy, J.E. Littlewood, G. P\`olya, Inequalities,
(2nd).Cambridge: Camb.Univ.Press, 1952, 44-45.

\REF{[7]} Albert W. Marshall, Olkin Ingram, Barry C. Arnold,
Inequalities: Theory of Majorization and Its Applications, (2nd).
New York, Dordrecht, Heidelberg, London: Springer, 2011, 8-10.

\end{document}